\documentclass[amsmath,amssymb,nofootinbib,aps]{revtex4}
\usepackage{graphicx}
\usepackage{dcolumn}
\usepackage{color}
\usepackage{bm}
\usepackage{amsmath,amssymb,graphicx,hyperref}
\usepackage{epsfig}
\usepackage{enumerate}
\usepackage{array}
\usepackage{multirow}

\begin{document}

\title{Entropy bounds, Geroch process, and the sign of deformation parameter}

\author{Bijan Bagchi$^1$\footnote{bbagchi123@gmail.com}, Akshat Pandey$^2$\footnote{apandey.physics@gmail.com}, Parmest Roy$^3$\footnote{pr556@snu.edu.in}, Sauvik Sen$^3$\footnote{sauviksen.physics@gmail.com} }
\affiliation{$^{1}$Department of Applied Mathematics, University of Calcutta, 92 Acharya Prafulla Chandra Road, Kolkata 700009, India\\
$^{2}$Department of Physics and Astronomy, KU Leuven,
Celestijnenlaan 200d, Leuven B-3001, Belgium\\
$^{3}$Department of Physics, Shiv Nadar Institution of Eminence, Gautam Buddha Nagar, Uttar Pradesh 203207, India}

\vskip-2.8cm
\date{\today}
\vskip-0.9cm

\begin{abstract}
Based on Geroch's process of dropping a system into a black hole from the vicinity of the
horizon, we investigate in this paper the influence of deformation on the Bekenstein entropy bound both for (3+1) and (2+1) dimensions in the context of a generalized uncertainty principle (GUP). While providing a coherent framework that sets an upper limit on the entropy across dimensions we show, within a semiclassical treatment, that while a negative GUP deformation yields a universal relaxation of the bound, a positive deformation tightens it. Our results may be interpreted as a
response to Planck-scale modifications of the near-horizon redshift.

\end{abstract}

\maketitle

\section{Introduction}

The Bekenstein entropy bound \cite{bekensteinbound}, which imposes an upper limit on the entropy of a given matter system, has been one of the key ideas that has influenced subsequent developments in information theory \cite{hayden}. This bound constrains the entropy function \(S\) to satisfy the inequality \(S \leq \lambda R E\) in a spatial region of size \(R\), where \(E\) is the energy, \(\lambda\) is a constant of \(O(1)\) and the right side is independent of \(G\), the universal gravitational constant. The upper limit is relevant in the context of the holographic principle, (see, for example, \cite{tho, suss1, suss2}). Subsequent work has revealed that black holes of the size of the Hubble horizon produce the maximal entropy inside the universe \cite{east,venz1,venz2,bak,kal,cai}. It turns out that for a given volume \(V\) with surface area \(A\), the entropy bound is given by \(S \leq A/(4G)\). More recently, Verlinde \cite{verl} gave a local version of it in terms of the Hubble parameter \(H\), namely \(S \leq (n-1)HV/(4G)\), where \(n\) represents the number of spatial dimensions of the universe. \\

In the literature, there have been numerous attempts to improve upon Bekenstein's bound. Within the framework of the generalized uncertainty principle (GUP), which accounts for a modification of the short-distance structure (see, for example, \cite{mag3,kempf,adler1,adler2, bagchi}) that is chiefly motivated to prevent arbitrarily sharp localization in space, thus addressing the minimal-length problem in quantum gravity and string theory (see, for example, \cite{Taw,rovelli,das,gross,amati,hossenfelder,lake}), thermodynamic arguments have been advanced to write down a modified wave-particle duality \cite{ahl}, and correction factors to the Bekenstein bound have been estimated \cite{buo}. Entropy bounds in scenarios with varying $G$ have also been explored, including cases that allow negative values of the GUP parameter \cite{ong, ongold}. 
Note that schemes with positive deformation are typically associated with the minimal length requirement leading to stronger entropy constraints, while negative ones can arise in alternative effective descriptions within relaxed entropy limits \cite{bizet}. \\

Adopting Planck units conforming to $c=\hbar=k=1$, a phenomenologically viable form of the GUP in (3+1) dimensions is typically given by \cite{mag3,kempf,adler1,adler2,bagchi}

\begin{equation}
    \Delta x \Delta p \geq \frac{1}{2} \left [ 1+ 4\beta \left (\frac{\Delta p}{M_{Pl,4}}\right)^{2}  \right ]
\end{equation}
where $\beta$ is the dimensionless GUP parameter and $M_{Pl,4}$
is the Planck mass in three space dimensions.
The relevance of the above corrected GUP to sub-Planckian black holes has aroused a great deal of interest \cite{bambi, carr1, carr2}. For a phenomenological treatment, this requires introducing an additional dimensionless constant to fix the lower bound of $\Delta x$.  
For its implementation, one of the criteria adopted is to have a Compton wavelength benchmark that would be approached by the size of the radius of curvature of spacetime at mass scales $\ll M_{Pl,4}$.\\

On the other hand, in the lower (2+1) dimensions, the following replacement of $\Delta x$ has been advocated \cite{iorio} 

\begin{equation}
    \Delta x \Delta p \geq \frac{1}{2} \left[ 1+ 4\gamma \left(\frac{\Delta p}{M_{Pl,3}}\right)^{3/2} \right]
\end{equation}
where $\gamma$ is the dimensionless GUP deformation parameter accounting for small corrections at the Planck scale corresponding to particle mass scales $\ll M_{Pl,3}$, where $M_{Pl,3}$ is the Planck mass in two space dimensions.
In terms of the reduced Compton wavelength $\lambda_c$, 
the following constraint on $\Delta x$ can be worked out

\begin{equation}
    \Delta x \geq \frac{1}{2} \left [\lambda_c + 4\gamma \frac{1}{\lambda_c^{\frac{1}{2}}} \left (\frac{1}{M_{Pl,3}}\right )^{\frac{3}{2}}  \right ].
\end{equation}
The first term on the right is one-half of the Compton wavelength $\lambda_c$, which is what one would have obtained from the standard Heisenberg uncertainty principle should the deformation term be turned off. Because in $(2+1)$ dimensions Newton's constant has inverse mass dimension,
the Planck quantities scale as $M_{Pl,3}\sim 1/G$ and $l_P\sim G$, where $G$
is the gravitational constant in two spatial dimensions. Dimensional consistency
of the GUP correction in Eq.~(2), relative to the leading Heisenberg term
$\Delta x\Delta p$, requires it to be written in terms of the dimensionless
ratio $\Delta p/M_{Pl,3}$. In the $(2+1)$-dimensional phenomenological scaling
adopted here, this ratio appears as $(\Delta p/M_{Pl,3})^{3/2}$, distinguishing
it from the usual quadratic correction in $(3+1)$ dimensions. At temperatures around the Planck mass scale, the quantum gravitational effects become non-negligible \cite{mukh, deeg, kot}, in that for the corresponding Planck energy, the gravitational force between two particles may turn out to be of the same order compared to any other operative force between them. \\

In the present work, we addressed two important but related issues by utilizing the Geroch process as nicely explained in \cite{boussoreview}. First, we inquired how the upper limit of the Bekenstein bound, originally derived for (3+1)-dimensional asymptotically flat spacetime, changed as the dimension is reduced by one spatial unit. Second, we investigated how its geometry got modified by minimal-length effects in both (3+1) and (2+1) dimensions, and then proceeded with examining the consequences thereof. Indeed, it demonstrated that the flat-space form of the bound survives in the AdS background and in lower dimensions too. Also, the implementation of Geroch's method provided us with three distinct advantages. First, we could 
show explicitly how the near-horizon redshift
removes the dependence at the AdS length scale to the leading semiclassical order. Second, we could envisage a common platform for comparing the
asymptotically flat Schwarzschild criterion with the asymptotic (2+1)
case. Third, we could make explicit the sign dependence of the GUP parameter. We found that a negative deformation somewhat eases the entropy bound, while a positive sign tightens it. 
Finally, the GUP-driven bounds derived below are not meant to replace the ordinary Bekenstein bound, but serve to test the stability under short-distance, quantum-gravity-motivated deformations. We may also remark here that in the perspective of Ong's recent work \cite{ong} on the significance of the sign of the deformation parameter appearing in the GUP-modified entropy bounds, our treatment has relied upon a modified wave-particle relation, and implemented the deformation issue through the effective mass/redshift structure.
\\

The organization of the paper is as follows. In Section~II, we introduce the GUP-corrected Arnowitt--Deser--Misner mass $M_{\rm ADM}$ \cite{bambi,ADM,vollick} for the $(3+1)$-dimensional Schwarzschild black hole and re-examine the resulting quantum-gravitational deviations from the classical Bekenstein bound. In Section~III, we present a self-contained derivation of the Bekenstein bound for a non-rotating $(2+1)$ dimensional black hole. In Section~IV, we extend our analysis to the GUP-corrected case by introducing an effective mass parameter $M_{\rm eff}$. Using the semiclassical result of Section~III, we derive the generalized entropy bounds for both signs of the GUP deformation parameter. In particular, we show how the correction can be expressed in terms of the system size $R$ after eliminating or optimizing over the auxiliary black-hole mass. Finally, Section~V contains a summary of our results.

\section{Impact of GUP corrections in $(3+1)$ dimensions}

In a general setting of $(3+1)$ dimensions, it is instructive to introduce the Arnowitt-Deser-Misner (ADM) mass as an effective way of encoding the GUP correction in terms of near-horizon thermodynamic quantities. It is given by \cite{carr1}

\begin{equation}
M_{ADM} = M + \frac{\beta}{2} f\left (\frac{M_{Pl,4}}{M} \right )
\label{eq:madm}
\end{equation}
where $M$ is the Schwarzschild mass, and $\beta$ is the GUP deformation parameter introduced in equation (1).

For brevity, we consider the simplest case of a linear representation of
$f$. For a Schwarzschild black hole, the near-horizon redshift factor for
a system whose centre of mass is at a proper distance $R$ from the horizon
is
\[
    \chi(R) \simeq \frac{R}{2R_s},
\]
where $R_s=2M/M_{\rm Pl,4}^2$ \cite{boussoreview}. Thus, a system of local energy
$E$ lowered to this point delivers the minimum energy
\[
    \delta M_{\rm min} \simeq E\chi(R)
    =\frac{ER}{2R_s}.
\]
Replacing $M\rightarrow M_{\rm ADM}$ in the Schwarzschild radius to extract
the influence of the GUP, we obtain

\begin{equation}
\delta M_{\rm min} \simeq
\frac{ERM_{Pl,4}^2}{4\left(M + \dfrac{\beta M_{Pl,4}^3}{2M}\right)}
=
\frac{ERM_{Pl,4}^2}{4}\,
\frac{M}{M^2+\dfrac{\beta M_{Pl,4}^3}{2}},
\qquad \left(M^2+\frac{\beta M_{Pl,4}^3}{2}>0\right).
\label{eq:linear_deltaM}
\end{equation}
Alongside, the change in the entropy is given by

\begin{equation}
S_{SCH}= \frac{A}{4G_4} = \frac{4 \pi}{M_{Pl,4}^2} \left(M + \frac{\beta M_{Pl,4}^3}{2M}\right)^2 \approx \frac{4 \pi M^2}{M_{Pl,4}^2} + 4 \pi \beta M_{Pl,4}
\label{14}
\end{equation}
which points to the ratio

\begin{equation}
\frac{\delta S_{\rm SCH}}{\delta M}
=
\frac{8\pi M}{M^2_{Pl,4}} .
\label{eq:linear_entropy_derivative}
\end{equation}
Clearly, it shows invariance when compared with the unmodified Schwarzschild case. For the optimally lowered system, the corresponding minimum increase of the black-hole entropy is
\[
\Delta S_{\rm SCH}^{\rm min}
=
\frac{dS_{\rm SCH}}{dM}\,\delta M_{\rm min}.
\]
Using Eqs.~\eqref{eq:linear_deltaM} and \eqref{eq:linear_entropy_derivative}, we obtain
\[
\Delta S_{\rm SCH}^{\rm min}
\simeq
\frac{8\pi M}{M_{Pl,4}^2}
\left[
\frac{ERM_{Pl,4}^2}{4}\,
\frac{M}{M^2+\dfrac{\beta M_{Pl,4}^3}{2}}
\right]
=
2\pi ER\,
\frac{M^2}{M^2+\dfrac{\beta M_{Pl,4}^3}{2}}.
\]
We observe that the generalized second law then requires that the entropy of the matter system that is dropped into the black hole must not exceed the corresponding minimum increase of the black-hole entropy. Thus, it transpires that
\begin{equation}
S_{\rm GUP} \le 2\pi ER\,
\frac{M^2}{M^2+\dfrac{\beta M_{Pl,4}^3}{2}},
\qquad \left(M^2+\frac{\beta M_{Pl,4}^3}{2}>0\right).
\label{eq:linear_mass_bound}
\end{equation}
where the condition in the parentheses follows from requiring the GUP-corrected effective ADM mass to remain positive. For a fixed black-hole mass $M$, this bound is valid for either sign of the GUP deformation parameter, provided the denominator remains positive. The pre-factor is smaller than unity for $\beta>0$, so the bound is tightened relative to the standard Bekenstein result. In contrast, for $\beta<0$, the prefactor is larger than unity, and the bound stays relaxed. \\

Equation~\eqref{eq:linear_mass_bound} does not yet look like a universal Bekenstein-type bound due to the dependence on $M$. To express the result entirely in terms of $R$, we impose the horizon-size condition
\[
R\leq R_s:=\frac{2M}{M_{Pl,4}^2},
\]
or equivalently
\[
M\geq M_R:=\frac{RM_{Pl,4}^2}{2}.
\]
The elimination of $M$ must be handled separately for the two signs of $\beta$ because of different monotonic behavior of the pre-factor for the two branches. \\

Consider the situation where $\beta<0$. Defining a new quantity
\[
h(M,\beta)=\frac{M^2}{M^2+\dfrac{\beta M_{Pl,4}^3}{2}}.
\]
we observe $h(M,\beta) > 1$  to be a decreasing function of $M$. Hence, using $M\geq M_R$ we run into the following inequality 
\[
h(M,\beta)\leq h(M_R,\beta).
\]
This translates to the upper limit 

\begin{equation}
S_{\rm matter} \le 2\pi ER\,
\frac{M^2}{M^2+\dfrac{\beta M_{Pl,4}^3}{2}}
\le
2\pi ER\,
\frac{M_R^2}{M_R^2+\dfrac{\beta M_{Pl,4}^3}{2}},
\qquad (\beta<0).
\end{equation}
Using $M_R=RM_{Pl,4}^2/2$, we are led to the result

\begin{equation}
S_{\rm matter} \le
2\pi ER\,
\frac{1}{1+\dfrac{2\beta}{R^2M_{Pl,4}}},
\qquad (\beta<0).
\end{equation}

For $\left|\dfrac{2\beta}{R^2M_{Pl,4}}\right|\ll1$, on carrying out a Taylor expansion we infer that
\begin{equation}
S_{\rm matter}\le 2\pi ER\left(1-\dfrac{2\beta}{R^2M_{Pl,4}}\right),
\qquad (\beta<0).
\end{equation}
For $\beta<0$, the above result is indicative of a positive correction to the Bekenstein bound.\\

For $\beta>0$, the quantity $h(M,\beta)<1$ and $h(M,\beta)$ is seen to increase with $M$. Therefore, from
\[
S_{\rm matter}\leq 2\pi ER h(M,\beta)
\]
one cannot directly replace $M$ by the lower value $M_R$. Instead, the universal bound is obtained by requiring the inequality to hold for a black-hole mass $M\geq M_R$. This gives

\begin{equation}
S_{\rm matter}\leq
\inf_{M\geq M_R}
\left[
2\pi ER
\frac{M^2}{M^2+\dfrac{\beta M_{Pl,4}^3}{2}}
\right].
\end{equation}
Since $h(M,\beta)$ increases with $M$ for $\beta>0$, the infimum is attained at the smallest allowed value, $M=M_R$. As such, we have the inequality

\begin{equation}
S_{\rm matter}\leq
2\pi ER
\frac{M_R^2}{M_R^2+\dfrac{\beta M_{Pl,4}^3}{2}}
\end{equation}
For $\left|\dfrac{2\beta}{R^2M_{Pl,4}}\right|\ll1$, it thus turns out
\begin{equation}
S_{\rm matter}\le 2\pi ER\left(1-\dfrac{2\beta}{R^2M_{Pl,4}}\right),
\qquad (\beta>0).
\end{equation}
Since the deformation parameter $\beta>0$, the correction is negative, and therefore the Bekenstein bound can be arbitrarily controlled.\\

Few remarks are in order.  Buoninfante et al \cite{buo} employed Ahluwalia's \cite{ahl} approach of deforming de-Broglie wavelength via GUP. By employing thermodynamic arguments they obtained a correction term to the next-to-leading order in the deformation parameter that reads

\begin{equation}
S \leq 2 \pi E R \left(1 + \frac{l_{P}^2|\beta|}{8  R^2}\right)
\label{15}
\end{equation}
They called it the generalized Bekenstein bound
and observed that for $\beta < 0$ the entropy of the system exceeded Bekenstein's limit but suppressed by $(\frac{l_{P}}{R})^2$ that forbade any plausible experimental test. Be that as it may, they also pointed out that Heisenberg's uncertainty principle may not also hold for negative deformation parameter.
We note that our entropy bound, apart from some inessential numerical factors, has a similar form as (\ref{15}).
Our derivation agrees qualitatively with that of Buoninfante et al. \cite{buo} and also with the sign-sensitive interpretation emphasized by Ong \cite{ong}: $\beta < 0$ relaxes the bound while $\beta>0$ tightens it. The distinction is that our correction is obtained through the Geroch-process implementation of the near-horizon redshift, rather than solely through a modified de-Broglie relation or by taking recourse to any thermodynamic argument. With the convention $R_s=2M/M_{Pl,4}^2$, the Geroch-process implementation gives a universal correction written solely in terms of $R$, with the leading term turning out to be proportional to $\beta/(R^2M_{Pl,4})$. \\

\section{Bekenstein bound in (2+1) dimensions: A quick derivation}

Interest in lower-dimensional gravity dates back to the foundational work of Ba\~{n}ados et al. (BTZ), who showed that a simplified formulation in $(2+1)$ dimensions admits a black-hole solution \cite{ban}. For a recent assessment of a BTZ black hole see \cite{bagchi1}. When expressed in terms of 

its angular momentum $J$, the BTZ metric in planar
polar coordinates reads \cite{carlip}

\begin{equation}
    ds^2 = - f(r)dt^2 + f(r)^{-1}dr^2 + r^2(d\phi - \frac{4G_3J}{r^2}dt)^2
    \label{1}
\end{equation}
involving the coordinate ranges are $0<r<\infty$, $-\infty<t<\infty$, $0 \leq \phi < 2\pi$, \(l\) denotes the AdS curvature radius and related to the corresponding cosmological constant through $\Lambda=-\frac{1}{l^2}$, and the metric function is given by \cite{maeda}
\begin{equation}
    f(r) = -8G_3M + \frac{r^2}{l^2} + \frac{(4G_3J)^2}{r^2} \xrightarrow{r\rightarrow \infty} \frac{r^2}{l^2}
    \label{2}
\end{equation} \\
where $M$ stands for the classical BTZ black-hole mass parameter appearing in the metric function.\\

The zeros of the metric function $f(r)$ characterize the event horizon. For the nonrotating case $(J=0)$, the horizon radius of the BTZ black hole is given by $r_{BTZ}=l\sqrt{8G_3M}$. Consequently, the expressions for the area $A_{\text{BTZ}}$ and entropy $S_{\text{BTZ}}$ acquire the forms 
\begin{equation}
    A_{\text{BTZ}} = 2 \pi r_{BTZ} = 2 \pi l\sqrt{8G_3M}; \quad S_{\text{BTZ}} = \frac{A}{4G_3} = \frac{\pi l}{2} \sqrt{\frac{8M}{G_3}}
    \label{3}
\end{equation}
In the following \((2+1)\)-dimensional discussion we write
\(G\equiv G_3\) for notational simplicity.\\

Casini \cite{casini} pointed out several years ago that the Bekenstein bound could be formulated so that its limiting value becomes independent of the number of spatial dimensions. To examine this point clearly, let us take recourse to Geroch's process of energy extraction \cite{geroch} in the ideal setup of a weakly gravitating, thermodynamically stable system, and conforming to the smallest sphere of radius $R$ that circumscribes the system. We bring such a system close to the horizon of a nonrotating BTZ black hole, whose center of mass is located at a distance less than $R$, and drop it into the black hole. This would produce a change in the mass resulting in an increase of entropy of the black hole. 

To calculate the energy acquiesced by the black hole,
we consider the redshift factor \(\Lambda(r)\) as defined by \cite{myers}
\begin{equation}
    \Lambda(r)\equiv \sqrt{f(r)} .
    \label{eq:redshift_factor}
\end{equation}
which gives 
\begin{equation}
    f(r)=-8GM+\frac{r^2}{l^2}
    =
    \frac{r^2-r_H^2}{l^2},
    \qquad
    r_H=l\sqrt{8GM}.
    \label{eq:btz_f_horizon}
\end{equation}
\\

We next introduce a near-horizon parameter \(\epsilon\) by setting
\begin{equation}
    r=r_H+\epsilon,
    \qquad
    \epsilon\ll r_H .
\end{equation}
To leading order near the horizon, we find
\begin{equation}
    f(r_H+\epsilon)
    =
    \frac{(r_H+\epsilon)^2-r_H^2}{l^2}
    \simeq
    \frac{2r_H\epsilon}{l^2}.
    \label{eq:btz_near_horizon_f}
\end{equation}
As a result, the redshift factor becomes
\begin{equation}
    \Lambda(r_H+\epsilon)
    \simeq
    \sqrt{\frac{2r_H\epsilon}{l^2}} .
    \label{eq:btz_near_horizon_redshift}
\end{equation}
Evaluation of the proper distance \(R\) from the horizon to \(r=r_H+\epsilon\) is elementary as depicted below
\begin{equation}
    R
    =
    \int_{r_H}^{r_H+\epsilon}\frac{dr}{\sqrt{f(r)}}
    \simeq
    \int_0^\epsilon
    \frac{d\epsilon'}{\sqrt{2r_H\epsilon'/l^2}}
    =
    \sqrt{\frac{2l^2\epsilon}{r_H}} .
    \label{eq:btz_proper_distance}
\end{equation}
Eliminating \(\epsilon\) between
\eqref{eq:btz_near_horizon_redshift} and
\eqref{eq:btz_proper_distance} gives
\begin{equation}
    \Lambda(R)
    \simeq
    \frac{r_H}{l^2}R
    =
    \frac{\sqrt{8GM}}{l}R .
    \label{eq:btz_redshift_R}
\end{equation}

In Geroch's process, the matter system is lowered adiabatically until its
centre of mass is at a proper distance of order $R$ from the horizon.
The energy delivered to the black hole, as measured at infinity, is the
local energy $E$ multiplied by the gravitational redshift factor. Therefore,
using Eq.~(25), the minimum energy delivered to the black hole is
\begin{equation}
    \delta M_{\rm min}
    \simeq E\Lambda(R)
    =
    \frac{ER\sqrt{8GM}}{l}.
\end{equation}
The corresponding increase of the black-hole entropy is
\begin{equation}
    \Delta S_{\rm BTZ}^{\rm min}
    =
    \frac{dS_{\rm BTZ}}{dM}\,\delta M_{\rm min}.
    \label{eq:btz_entropy_increase_general}
\end{equation}
Using the connection
\begin{equation}
    S_{\rm BTZ}
    =
    \frac{\pi l}{2}\sqrt{\frac{8M}{G}},
\end{equation}
we obtain
\begin{equation}
    \frac{dS_{\rm BTZ}}{dM}
    =
    \frac{2\pi l}{G}\sqrt{\frac{G}{8M}} .
    \label{eq:btz_entropy_derivative}
\end{equation}
This implies
\begin{equation}
    \Delta S_{\rm BTZ}^{\rm min}
    =
    \frac{2\pi l}{G}\sqrt{\frac{G}{8M}}\,
    \frac{ER\sqrt{8GM}}{l}
    =
    2\pi ER .
    \label{eq:btz_entropy_increase_final}
\end{equation}
Since the generalized second law of thermodynamics requires that the entropy of the dropped matter cannot exceed the above value, we conclude

\begin{equation}
    S_{\rm matter}\leq 2\pi ER .
    \label{eq:btz_bekenstein_bound}
\end{equation}
Interestingly, this upper limit has the same form as the Bekenstein bound obtained in
\((3+1)\)-dimensional asymptotically flat spacetime. Here, however, the
derivation uses the near-horizon region of the non-rotating BTZ black hole.
Although the intermediate expressions depend on the AdS scale \(l\), this
dependence cancels in the final expression of the entropy bound. Having established the semiclassical BTZ result, we next estimate how it is modified when phenomenological GUP corrections are taken into account

\section{Influence of GUP}

In the presence of GUP, we encode the leading correction in the BTZ
mass parameter through the effective replacement
\begin{equation}
M_{\rm eff}
=
M+\frac{\beta}{2}\left(\frac{M_{{\rm Pl},3}}{M}\right)^{3/2},
\tag{32}
\end{equation}
where \(M_{{\rm Pl},3}\) is the Planck mass in two spatial dimensions.
This convention is the reason why the gravitational constant appears explicitly in the BTZ
intermediate formulae, whereas in the \((3+1)\)-dimensional discussion
it is mostly hidden inside \(M_{{\rm Pl},4}\).

Within the same effective semiclassical prescription, the near-horizon
redshift factor is obtained from the BTZ expression by replacing
\(M\rightarrow M_{\rm eff}\). Hence, for a system of energy \(E\) and
proper size \(R\),
\begin{equation}
\delta M_{\rm min} \simeq E\Lambda(R)
=
ER\,\frac{\sqrt{8G M_{\rm eff}}}{l}.
\tag{33}
\end{equation}
The corresponding BTZ entropy is evaluated consistently with the same
effective replacement in the semiclassical expression,
\begin{equation}
S_{\rm BTZ}
=
\frac{\pi l}{2}\sqrt{\frac{8M_{\rm eff}}{G}} .
\tag{34}
\end{equation}
Differentiating with respect to the original BTZ mass parameter \(M\)
gives
\begin{equation}
\frac{dS_{\rm BTZ}}{dM}
=
\frac{\pi l}{4}
\sqrt{\frac{8}{G M_{\rm eff}}}\,
\frac{dM_{\rm eff}}{dM}.
\tag{35}
\end{equation}
Combining Eqs.~(33) and (35), we obtain
\begin{equation}
\Delta S_{\rm BTZ}^{\rm min}
=
\frac{dS_{\rm BTZ}}{dM}\,\delta M_{\rm min}
\simeq
2\pi ER\,\frac{dM_{\rm eff}}{dM}.
\tag{36}
\end{equation}
The generalized second law then requires
\(S_{\rm matter}\leq \Delta S_{\rm BTZ}^{\rm min}\). Therefore,
\begin{equation}
S_{\rm matter}
\leq
2\pi ER\,\frac{dM_{\rm eff}}{dM}
=
2\pi ER
\left(
1-\frac{3\beta M_{{\rm Pl},3}^{3/2}}{4M^{5/2}}
\right).
\tag{37}
\end{equation}
This expression reduces to the standard Bekenstein bound
\(S_{\rm matter}\leq 2\pi ER\) when \(\beta\to 0\). For fixed BTZ mass
\(M\), positive deformation tightens the bound, while negative
deformation relaxes it. Thus the appearance of \(G\) in the
\((2+1)\)-dimensional derivation is not an inconsistency; it enters
through the BTZ redshift and entropy formulae and cancels from the
undeformed final bound, just as shown in Eq.~(31).\\

\begin{figure}
    \centering
    \includegraphics[width=0.5\linewidth]{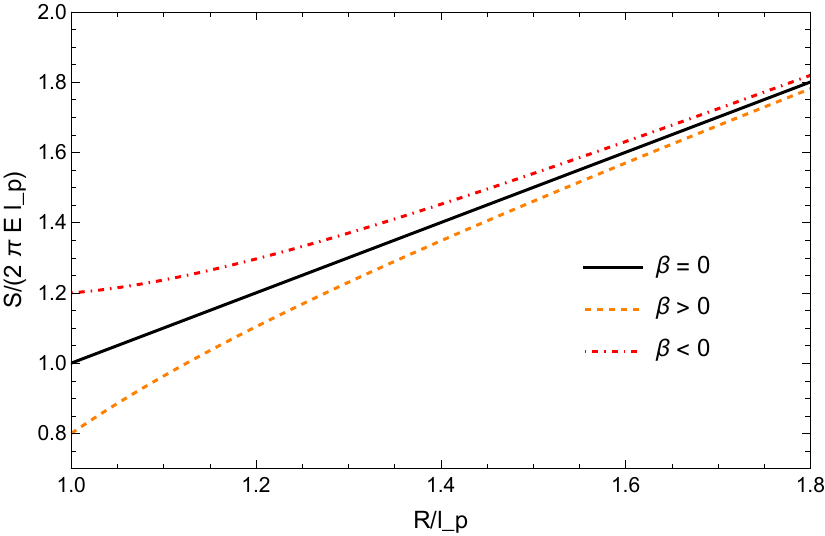}
    \caption{
GUP-corrected Bekenstein entropy bound in $(2+1)$ dimensions, plotted in terms of the dimensionless variables $x=R/l_{P}$ and $y=S/(2\pi E l_{P})$. 
The solid black line denotes the standard Bekenstein bound $y=x$. 
The dashed orange curve corresponds to positive deformation parameter $\beta>0$, for which the entropy bound is tightened, while the dot-dashed red curve corresponds to negative deformation parameter $\beta<0$, for which the bound is relaxed. 
For visualization, we use the leading-order form 
$y=x\left(1\mp \epsilon_{2+1}/x^5\right)$ 
with effective dimensionless correction strength $\epsilon_{2+1}=0.2$. 
The plot is shown over the restricted range $1\le x\le 1.8$ in order to make the GUP-induced deviation visible. This zoomed-in presentation is physically motivated, since in $(2+1)$ dimensions the correction falls rapidly with increasing $R$, implying that the deviation from the classical bound scales as $R^{-4}$ and becomes negligible away from the Planck-scale region.
}
    \label{fig:gup21}
\end{figure}

As a final comment, let us note that the correction in the $(2+1)$-dimensional case is strongly suppressed for systems whose size is large compared with the Planck scale. In the dimensionless form used in Fig.~\ref{fig:gup21}, the correction falls as $R^{-4}$, so that the GUP-induced deviation is appreciable only near the Planck regime. This is consistent with the interpretation of the GUP correction as a short-distance quantum-gravitational effect.

\section{Summary}

In this paper, adopting Geroch's method of adding a mass to a non-rotating black hole and thus influencing the growth of its horizon area, we looked into the modification of Bekenstein bound that is brought about. In the presence of GUP, we first derived a general upper bound for the entropy in (3+1) dimensions and then transformed it to the Bekenstein-type bound by making use of the horizon-type condition. We also discussed the specific cases of the sign parameter associated with the deformation and compared our results with the existing works in the literature. Proceeding next to the case of (2+1) dimensions, we demonstrated that the usual upper bound
\(S_{\rm matter}\leq 2\pi ER\) could be recovered, and that how it got modified when the GUP effects were encoded through an
effective mass parameter entering in the near-horizon redshift. Within a semiclassical prescription, the
sign of the deformation parameter was seen to control the direction of the correction:
positive deformation tightened the bound, whereas negative deformation
eased it. In this way, the GUP-corrected bounds were found to indicate the sensitivity to
short-distance quantum-gravitational modifications. Since the corrections were suppressed by inverse powers of the system size in Planck units, the standard Bekenstein bound could be recovered as we moved away from the Planck regime. Our corrected bounds may be interpreted as leading-order effective results which are valid only in the
perturbative regime where the GUP correction is small and the entropy bounds remain positive.\\

\section{Acknowledgment}

We thank the referee for many constructive comments.

\end{document}